# Grating magneto-optical trap of cesium atoms with an additional retroreflected laser beam

Akifumi Takamizawa[1,a)], Ryohei Hokari[2], Sota Kagami[3,4], Thu H. H. Le[2], Ryohei Takei[5], Ken Hagimoto[1], and Shinya Yanagimachi[1]

[1]*National Metrology Institute of Japan, National Institute of Advanced Industrial Science and Technology (AIST), 1-1-1 Umezono, Tsukuba, Ibaraki 305-8563, Japan*

[2]*Advanced Manufacturing Research Institute, National Institute of Advanced Industrial Science and Technology (AIST), AIST Tsukuba East 1-2-1 Namiki, Tsukuba, Ibaraki 305-8564 Japan*

[3]*Secure System Platform Research Laboratories, NEC Corporation, 1753 Shimonumabe, Nakahara-ku, Kawasaki, Kanagawa 211-0011, Japan*

[4]*National Institute of Advanced Industrial Science and Technology, NEC-AIST Quantum Technology Cooperative Research Laboratory, 1-1-1 Umezono, Tsukuba, Ibaraki 305-8568, Japan*

[5]*Device Technology Research Institute, National Institute of Advanced Industrial Science and Technology, Ibaraki 305-8564, Japan*

[a)]Authors to whom correspondence should be addressed: akifumi.takamizawa@aist.go.jp

**Abstract**

A magneto-optical trap of cesium atoms was generated by applying a circularly polarized cooling laser beam onto a reflective two-dimensional diffraction grating with an aperture and by retroreflecting the incident beam passing through the aperture while reversing the circular polarization. The cooling laser beams comprised the incident, retroreflected, and four diagonally diffracted beams at an angle of 50°. The intensity of the retroreflected beam was carefully adjusted to balance the radiation forces acting on the atoms. Despite the challenges posed by cesium atoms with high nuclear spin, a significant number of cold atoms ($7.0 \times 10^6$) were captured when the detuning and power of the incident beam were −10 MHz and 131 mW, respectively, with the intensity of the retroreflected beam set to 69 % of that of the incident beam. The importance of the retroreflected beam in the trapping process was highlighted when the intensity ratio was reduced to 24 %, resulting in the absence of trapped atoms. This underscores the significance of the retroreflected beam in the trapping process. Notably, the distribution of the cold atom cloud differed from other magneto-optical traps, as it was not centered in the region where the cooling beams overlapped. Instead, numerous cold atoms were observed when the cloud was positioned near the apex closer to the grating side and an edge line of the overlapping region. Therefore, trapping can be achieved with the assistance of attractive dipole forces exerted by the diffracted beams, which exhibits high intensities at these positions.

Laser cooling of neutral atoms has resulted in ultra-high precision and accurate measurements in atomic clocks,[1-6] magnetometers,[7,8] and gravimeters.[9,10] This advancement is made possible by the extended interaction time with resonant lasers and minimized interference from the environment. Furthermore, laser-cooled atoms have proven to be valuable as qubits in quantum computation, leveraging their long coherence time, room-temperature operation, and high degree of freedom for the gate operation.[11-14]

In many instances, atoms cooled to the Doppler limit of approximately 100 μK are initially prepared using a magneto-optical trap (MOT).[15] Conventional MOTs utilize six cooling laser beams with diameters of approximately 10 mm, illuminating a region under an ultrahigh vacuum. Recent developments have focused on miniaturizing the MOT setup to enhance its portability for various applications. To downsize the optical components for MOT, optical integrated circuits and metasurfaces have been employed[16-18]. Moreover, vacuum systems, including cells and pumps, have been miniaturized.[19-23]

Despite advancements in shrinking optical and vacuum components, compactifying the setup remains challenging owing to the requirement of three-dimensional (3D) irradiation of thick laser beams. To address this issue, researchers have explored the use of a pyramidal reflector in an MOT setup that utilizes only one incident cooling beam.[24-26] However, this approach presents a drawback as the reflector must be placed in a vacuum environment to generate the MOT within the pyramidal shape. A novel configuration has been developed for an MOT that does not require optical components to be placed in a vacuum. This innovative MOT utilizes a planar diffraction grating, where the cooling laser beams comprise one incident and three or more reflectively diffracted beams.[17,22,27-32] In the grating MOT (gMOT), the cooling beams are oriented at nonorthogonal angles to each other,[33] creating a unique configuration, such as the tetrahedral or quadrangular pyramid. The gMOT provided cold $^{87}$Rb atoms totaling $6 \times 10^7$, which were further cooled to a temperature of 60 μK through post-cooling techniques.[28] Except for $^{87}$Rb, gMOTs have also been successfully demonstrated for other alkali metals $^{85}$Rb[22] and $^{7}$Li,[31] as well as alkali earth metals $^{87}$Sr[32] and $^{88}$Sr.[30] However, to the best of our knowledge, gMOTs of $^{133}$Cs (Cs hereinafter) atoms, which belong to alkali metal and a commonly treated atomic species in laser cooling, have not been achieved. The MOTs of Cs atoms have been successfully utilized, particularly in atomic clocks because the second is defined in the International System of Units by their transition frequency.[1] In quantum computing applications with cold Cs atoms, their large mass and high polarizability result in small recoil energy and deep trapping potential, respectively.[12,14]

While the energy level structures are consistent among alkali metals, the nuclear spin quantum number varies by atomic species. Consequently, the quantum numbers of the total atomic angular momentums of the ground states (excited states), $F$ ($F$'), differ among alkali metals. Theoretically, the realization of an MOT with nonorthogonal cooling beams is challenging unless the frequencies of all

σ− transitions between ground and excited states are negatively shifted by a magnetic field.[34] For the cooling transition in the D2 line of alkali metal atoms, where $F' = F + 1$, the requirement for the frequency shifts is satisfied when $F < 3$. [7]Li and [87]Rb ($F = 2$) satisfy this requirement, and [85]Rb ($F = 3$) lies on the boundary of the inequation. In contrast, the requirements for Cs are unsatisfactory ($F = 4$). Recent numerical calculations for a gMOT have shown that as $F$ increases, the radiation forces exerted on atoms by diagonal diffracted beams weaken in the direction perpendicular to the grating surface.[32] In our experiment involving Cs atoms in gMOT, no cold atoms were present when a 2D grating was utilized; however, when 1D gratings were utilized, cold atoms of $\ll 10^6$ were captured. The intensity per diffracted beam in the latter scenario was twice as high as that in the former case. These findings, along with previous theoretical analyses, suggest that the radiation forces acting opposite to the incident beam were insufficient for Cs atoms.

This study elucidates the generation of a gMOT of Cs atoms by introducing a cooling beam opposite to the incident beam to compensate for inadequate radiation forces. A reflective 2D grating with an aperture smaller than the incident beam diameter was employed. The incident beam passing through the aperture was retroreflected, whereas the incident beam illuminating the grating surface was diffracted diagonally. The incident and retroreflected beams formed a σ+-σ− polarization configuration. Moreover, the intensity of the retroreflected beam was adjusted to achieve a balance in radiation forces. Through this balanced gMOT setup, cold Cs atoms of $7.0 \times 10^6$ were captured.

The experimental setup of the balanced gMOT is shown in Fig. 1(a). The 2D grating with a dot-shaped microstructure was a square with sides of 66 mm and featured a square aperture with a side length of $2a = 10.0$ mm at the center, as shown in Fig. 1(b). The grating was positioned face-down on a quartz vacuum glass cell with a refractive index of 1.45 at the resonant wavelength of Cs D2 line, 852 nm. Both sides of the glass plates of the cell were anti-reflection-coated with a reflectivity of <1 % for incident angles of 0°–45°. The thickness of the two glass plates beneath the grating was 3.0 mm, with an inner cell height of 13.0 mm.

The cooling laser was slightly red-detuned from the transition between $6^2S_{1/2}$, $F = 4$ and $6^2P_{3/2}$, $F' = 5$ by $\Delta$ in terms of angular frequency, whereas the repump laser was tuned to the transition between $6^2S_{1/2}$, $F = 3$ and $6^2P_{3/2}$, $F' = 4$. Both lasers were supplied by external-cavity diode lasers. Following the amplification of the cooling laser power with a taper amplifier, the repump laser was overlaid onto the cooling laser through a polarization beam splitter. Both lasers were coupled to a common polarization-maintaining single-mode fiber and directed close to the glass cell. The output from the fiber was collimated to a beam with a diameter of $2w_1 = 19$ mm (full width at $1/e^2$ maximum intensity), and circular polarization was achieved using a quarter-wave plate (QWP). The beam then perpendicularly illuminated the grating surface. The powers of the incident beams of the cooling and repump lasers were $P_1 = 131$ mW at the maximum and 8.7 mW, respectively. The grating diffracted the incident beam in four directions at an angle of $\theta = 50°$ with respect to the normal direction of the grating surface as the first-order diffraction light. The diffraction efficiency $D_g$ was 86.9 %. The degree of the circular polarization of the diffracted beams was evaluated to be 86.8 % when the incident beam had complete circular polarization. The incident beam passed through the aperture of the grating and was retroreflected by a mirror. Moreover, a QWP and neutral density filter (NDF) with transmittance $T_{ND}$ were inserted on the path of the counterpropagating beams. The QWP was used to create σ⁺-σ⁻ configuration, whereas the NDF adjusted the intensity of the retroreflected beam. The counterpropagating beam intensity ratio (CBIR) was defined as the intensity ratio of the retroreflected to the incident beam, expressed as $\alpha = T_{ND}^2$, neglecting residual losses at the glass cell, mirror and QWP.

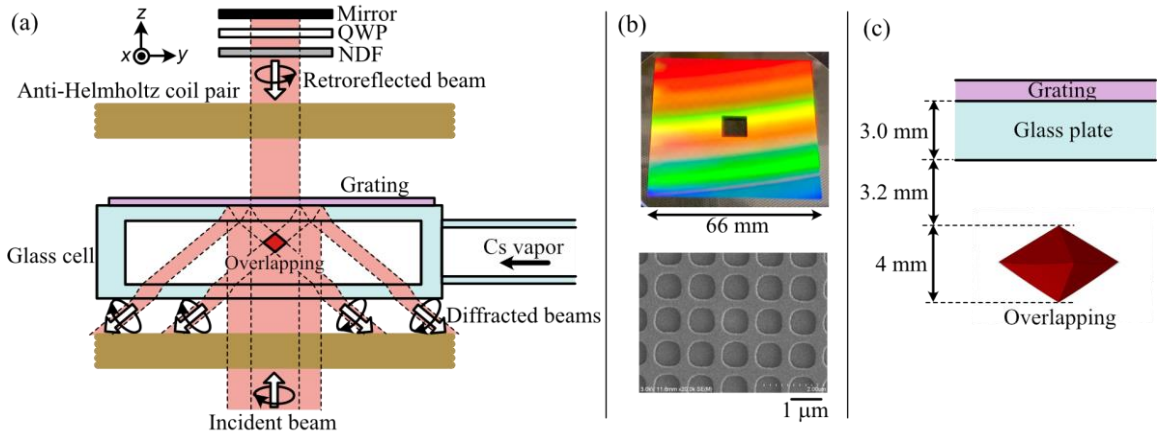

Fig. 1 (a) Side view of the experimental setup of the balanced gMOT. The orthogonal coordinate system is also indicated. The sides of the grating are parallel to the *x*- or *y*-axis. (b) (Top) Photograph of the 2D grating with an aperture. (Bottom) Electron microscopy image of the dot-shaped microstructure on the 2D grating. (c) Three-dimensional view of the overlapping region from the horizontal direction at an angle of 30° to the *x*-axis.

The shape and relative position of the region in which the cooling beams overlap are shown in Fig. 1(c). The volume and height of the overlapping region were calculated as 25 mm³ and 4 mm, respectively, with the top of the overlapping region positioned 3.2 mm below the inner surface of the cell glass plate.

Under a vacuum pressure of $7 \times 10^{-7}$ Pa, cesium vapor was introduced into the glass cell from a dispenser by applying an electric current for heating. By applying a quadrupole magnetic field with a gradient of approximately 0.1 T/m using an anti-Helmholtz coil pair, a cold atom cloud was generated near the center of the field. The position of the coil pair was adjusted using three-axis screws, where the two coils were unified to achieve parallel displacement of the magnetic field. Observation of the fluorescence emitted by the cold atom cloud was conducted by both a complementary metal oxide semiconductor (CMOS) camera and photodiode. The CMOS camera was utilized for imaging purposes, whereas the photodiode was employed to quantitatively measure the fluorescence power to determine the number of cold atoms, denoted as $N_a$ (refer to Section 1 of the supplementary material).

In most MOTs, including conventional gMOTs, a cold atom cloud is typically generated near the center of the overlapping region by adjusting the position of the quadrupole magnetic field center. However, in this experiment, the cold atom cloud was generated exclusively near the upper edge lines of the overlapping region. Even when the center of the quadrupole magnetic field was positioned within the overlapping region, the cloud dispersed into parts near the four upper edge lines, as shown in Fig. 2(b). The maximum number of cold atoms was achieved when the cloud was generated near the top of the overlapping region and close to one of the edge lines, referred to as the "optimum position" below, (Fig. 2(a)). As the CBIR increased, the cold atom cloud was pushed downward, prompting adjustments to the coil pair's position to elevate the cloud back to the optimum position. The fluorescence images of the cold atom clouds located at the optimum position at various CBIRs are shown in Figs. 2(c-f). The elongated cloud along the edge line expanded as the CBIR decreased.

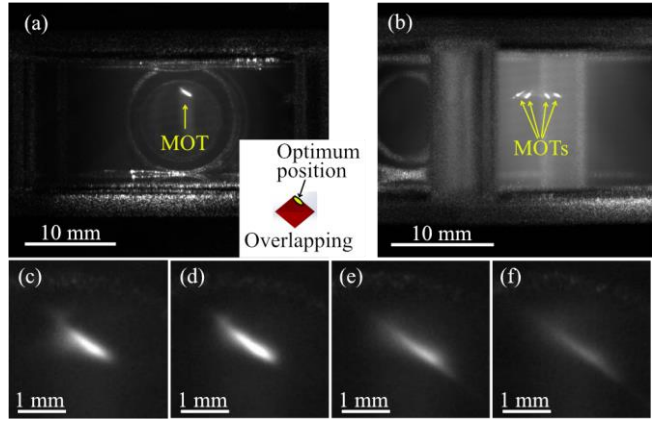

Fig. 2. (a) and (c-f) Fluorescence images of cold atom clouds at the optimum position. $\alpha$ = (c) 1.00, (d) 0.69, (e) 0.47, and (f) 0.40. The optimum position is shown at the bottom right of (a). (b) Fluorescence image with the cloud divided into four by moving the coil pair down by 1 mm. While the image in (b) is viewed laterally at an angle of 20° to the $y$-axis, the other images are viewed in the $y$-direction. Among (c-f), the settings of the CMOS camera were consistent to aid the comparison of the fluorescence intensities among the images. The fluorescence intensities in (a) and (b) cannot be compared with the other images because the camera settings were altered.

When $\Delta/(2\pi) = -10$ MHz, $P_1 = 131$ mW, and $\alpha = 0.69$, the number of cold atoms $N_a$ reached a maximum value of $7.0(3) \times 10^6$. The increase $N_a$ with loading time $t_L$ is shown in Fig. 3(a). The behavior of the number of cold atoms was consistent with the function $1 - \exp(-t_L/\tau)$, with a time constant $\tau = 0.15$ s. A graph of $N_a$ versus $-\Delta/(2\pi)$ at various $P_1$ is shown in Fig. 3(b), where $\alpha = 0.69$. The optimum frequency detuning $|\Delta|$ was approximately $2\Gamma$, where $\Gamma$ $(:= 2\pi \times 5.23$ MHz$)^{35}$ represents the natural linewidth of the excited state. The detuning shifted toward larger $|\Delta|$ as $P_1$ increased, similar to conventional MOTs. The graph $N_a$ versus $P_1$ is shown in Fig. 3(c), where $\Delta/(2\pi) = -10$ MHz and $\alpha = 0.69$. $N_a$ increased with increasing $P_1$ and was not saturated, indicating that enhancing the power of the cooling beams could further increase $N_a$.

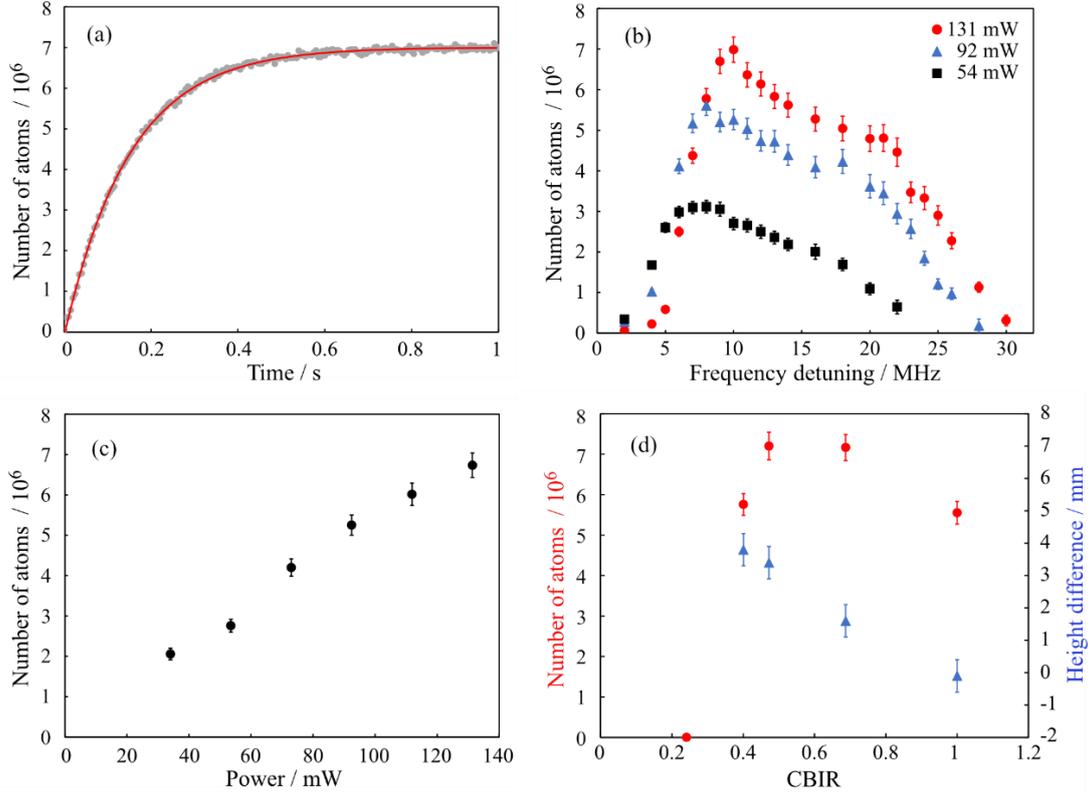

Fig. 3. Number of cold atoms $N_a$ as functions of (a) $t_L$, (b) $-\Delta/(2\pi)$, (c) $P_1$, and (d) $\alpha$. In (a), the gray dots and red line denote the experimental data and fitting curve to $N_a[1 - \exp(-t_L/\tau)]$, respectively. In (b), the red dots, blue triangles, and black squares represent the cases in which $P_1 = 131$ mW, 92 mW, and 54 mW, respectively. In (d), the red dots and blue triangles express $N_a$ (left axis) and $\delta h_a$ (right axis), respectively. Other than the changing parameters, the conditions were fixed at $\Delta/(2\pi) = -10$ MHz, $P_1 = 131$ mW, and $\alpha = 0.69$, with the error bars indicating the measurement uncertainties.

The graph of $N_a$ versus $\alpha$ is shown in Fig. 3(d), where $\Delta/(2\pi) = -10$ MHz and $P_1 = 131$ mW. Additionally, the height difference $\delta h_a \equiv h_a - h_f$ was plotted, where $h_a$ and $h_f$ denote the heights of the center of the cold atom cloud and quadrupole magnetic field, respectively. The center of the field was estimated based on the coil pair's position. An increase in $\delta h_a$ indicates that the coil pair was adjusted downward to bring the cold atom cloud to the optimum position. As shown in Fig. 3(d), when $\alpha = 1.00$, $\delta h_a = -0.1(5)$ mm. This indicates that the height of the cold atom cloud aligned closely with the center of the quadrupole magnetic field. As $\alpha$ decreased, the height difference $\delta h_a$ increased steadily. For $\alpha$ values of 0.47 and 0.69, corresponding to height differences of +3.4(5) and +1.6(5) mm, respectively, $N_a$ was approximately 30 % higher compared with when $\alpha = 1.00$. However, $N_a$ significantly decreased when $\alpha$ was reduced to 0.40. Further reduction to 0.24 resulted in the absence of cold atom clouds, despite the gradual lowering of the center of the quadrupole magnetic field to 11 mm below the optimal position. The value of $\alpha = 0.69$ was deemed optimal over 0.47 owing to the rapid decline in $N_a$ when $\alpha$ was decreased from 0.47.

Regarding the balance of the radiation forces, when $\delta h_a = 0$, the radiation forces on an atom at rest and positioned at the center of the quadrupole magnetic field are in equilibrium. In a simplified scenario where an atom transitions from a ground state $F = 0$ to an excited state $F' = 1$, the radiation force exerted by the cooling beam $i$, with saturation intensity, is proportional to the ratio of the intensity of beam $i$ to the total intensities of all cooling beams, $I_i/\sum_m I_m$ (refer to Section 2 of the supplementary material). Therefore, the balancing condition for the $z$ components of the radiation forces can be expressed as $I_1 - I_2 - \sum_{i=3}^{6} I_i \cos\theta = 0$, where $i = 1$ for the incident beam, $i = 2$ for the retroreflected beam, and $i = 3, 4, 5,$ and 6 for each diffracted beam. Substituting $I_2 = \alpha I_1$ and the numerical values of $I_i$ ($i \neq 2$) and $\theta$ in the experiment, we determined that the balancing condition will be satisfied when $\alpha$ is approximately 0.55. This is deviated from $\alpha = 1.00$, resulting in $\delta h_a = 0$ in the experiment.

Numerical calculations for the conventional gMOT with no retroreflected beams in Ref. 32, considering all Zeeman sublevels for an atom with $F > 0$, revealed that the position of a cold atom cloud shifted more significantly toward the grating from the center of the quadrupole magnetic field as $F$ increases (refer to Section 3 of the supplementary material). Applying this knowledge to a balanced gMOT indicates that, for an atom with a larger $F$, the retroreflected beam must be more intense to balance the $z$-component of the radiation forces. The discrepancy in balancing conditions between the theoretical analysis using a simple model with $F = 0$ and the experimental results for Cs atoms with $F = 4$ can be attributed to this consideration. Although numerical calculations for the balanced gMOT considering all Zeeman sublevels have not been conducted, the CBIR can be experimentally optimized.

Moving on to the location of the cold atom cloud, the intensity of individual diffracted beams was high on the upper slope surface of the overlapping region. This resulted from the edges of the aperture of the grating coinciding with areas of significant intensity change in the incident beam with a Gaussian distribution. Therefore, the diffracted beams with red detuning formed trapping potentials through attractive dipole forces near the slope surfaces. When $\Delta/(2\pi) = -10$ MHz and $P_1 = 131$ mW, the depth of the potential generated by one of the diffracted beams is calculated as 110 μK, approximately equal to the temperature at the doppler limit of Cs atoms. Moreover, the intensity of the red-detuned laser was enhanced owing to the superposition of the beams, resulting in a deeper dipole force potential at the optimum position. This potential can assist gMOT in confining cold atoms. Here, the number of cold atoms does not increase if the cloud size is too small, as the density of cold atoms is limited to approximately $10^{11}$ cm$^{-3}$ in an MOT.[36] The strength of confinement impacts the maximum value of $N_a$ when $\alpha = 1.00$ ($\delta h_a = -0.1$ mm) while the cloud was moderately spread out when $\alpha \approx 0.6$ ($\delta h_a \approx +2$ mm). However, the requirement of dipole force assistance for the gMOT of Cs atoms in contrast to other alkali metal atoms such as $^{87}$Rb remains unclear.

In conclusion, despite the challenges posed by Cs atoms' high nuclear spin, the successful implementation of a 2D grating with an aperture and a retroreflected cooling beam with adjusted intensity has enabled the realization of the gMOT. The integration of the grating with additional optical components, such as a mirror, QWP and NDF will simplify the optical setup for the balanced gMOT to be on par with that of a conventional gMOT. The balanced gMOT paves the way for the application to other atomic species with high nuclear spins, facilitating the miniaturization of apparatuses in this field.

**Supplementary Material**

1. Estimation of the number of cold atoms

The power of the fluorescence emitted from a single cold atom can be calculated as[35,37]

$$p_s = \frac{\hbar \omega s_0 \Gamma}{2[1+s_0+(2\Delta/\Gamma)^2]}, \quad (1)$$

where $s_0 = (\sum_i I_i)/I_s$. $I_i$ denotes the intensity of the cooling beam $i$ at the location of the atom and $I_s$ (:= 2.71 mW/cm$^2$)[35] represents the saturation intensity for isotropic light polarization. We considered $i = 1$ for the incident beam, $i = 2$ for the retroreflected beam, and $i = 3, 4, 5,$ and $6$ for the diffracted beams. Moreover, $\omega$, $\hbar$, and $\Gamma$ (:= $2\pi \times 5.23$ MHz)[35] represent the angular frequency of the cooling laser, Dirac constant, and natural linewidth of the excited states, respectively. Using $p_s$ in Eq. (1) and the fluorescence power detected by the photodiode, $p_{PD}$, the number of cold atoms was estimated as follows:

$$N_a = \frac{4\pi}{\Omega} \cdot \frac{p_{PD}}{p_s}, \quad (2)$$

assuming isotropic radiation, where $\Omega$ (:= $4\pi \times 1.2 \times 10^{-3}$) represents the solid angle of the fluorescence collected by a lens. This estimation considers changes in $p_s$ associated with changes in the experimental parameters $\Delta$, $P_1$, and $\alpha$.

The calculations for $I_i$ are outlined below. The intensity of the incident beam at the radial coordinate $r$ is given as $I_{\text{inc}}(r) = 2P_1(\pi w_1^2)^{-1}\exp[-2(r/w_1)^2]$. Neglecting the residual loss at the glass cell, QWP, and mirror, the intensity of the retroreflected beam is expressed as $I_{\text{ret}}(r) \simeq \alpha I_{\text{inc}}(r)$, whereas the intensity of one of the diffracted beams on the grating surface is expressed as $I_{\text{dif}}(r) \simeq [D_g/(n_d \cos\theta)]I_{\text{inc}}(r)$, with $n_d$ representing the number of diffracted beams. The division by $\cos\theta$ results from the thinning of the beams owing to diagonal diffraction. The beam intensities at the location of the cold atom cloud were derived from $I_{\text{inc}}(r)$, $I_{\text{ret}}(r)$, and $I_{\text{dif}}(r)$.

2. Radiation forces resulting from individual cooling beams in the simple case

When $F = 0$ and $F' = 1$, the Zeeman sublevel of the ground level is only $m_F = 0$, and the Clebsch–Gordan (CG) coefficients for $\sigma^+$, $\sigma^-$, and $\pi$ transitions are all unity. The radiation force exerted on an atom by cooling beam $i$ is expressed as[38]

$$\vec{F}_i = \hbar \vec{k}_i \frac{\Gamma}{2} \frac{I_i}{I_s} \sum_{j=-1,0,1} \frac{\eta_j}{1+s_0+\frac{4(\Delta-\vec{k}_i\cdot\vec{v}-\mu_F j|\vec{B}|)^2}{\Gamma^2}}, \quad (3)$$

where $\eta_{-1}$, $\eta_0$, and $\eta_{+1}$ denote the $\sigma^-$, $\pi$, and $\sigma^+$ polarization components of the cooling beam, respectively. $\vec{k}_i$, $\vec{v}$, $\mu_F$, and $\vec{B}$ denote the wavevector of the cooling beam $i$, velocity of the atom, Bohr magneton, and magnetic flux density, respectively. By utilizing Eq. (3), we observed that a stationary atom positioned at the center of a quadrupole magnetic field, where $\vec{v} = \vec{B} = 0$, experiences a radiation force $\vec{F}_i \simeq (\hbar \vec{k}_i \Gamma/2)(I_i/\sum_m I_m)$ when the laser intensities are saturated. That is $s_0 \gg 1$ and $(2\Delta/\Gamma)^2$.

3. gMOT of actual alkali metal atoms

The impact of the quantum number $F$ on the conventional gMOT is analyzed in detail in Ref. 32. Based on the study, we briefly discuss the challenges associated with the conventional gMOT of alkali metal atoms with high $F$ values, as well as highlight the benefits of utilizing a balanced gMOT. For the cooling transition in the D2 lines of alkali metal atoms, the ground level $F \geq 1$, and the excited level $F' = F + 1$. To accurately evaluate the radiation force acting on an actual alkali metal atom, the populations and CG coefficients must be considered, accounting for all the Zeeman sublevels $-F \leq m_F \leq F$ and $-F' \leq m_{F'} \leq F'$ for the ground and excited levels, respectively. In our analysis, we establish the center of the quadrupole magnetic field as the origin of the orthogonal coordinate system. In the region where $z > 0$ along the $z$-axis, the incident and retroreflected beams have polarizations of $\sigma^+$ and $\sigma^-$, respectively. Conversely, the diagonally diffracted beam has polarizations comprising $\sigma^+$ (67 %), $\pi$ (29 %), and $\sigma^-$ (3 %) when $\theta = 50°$. These proportions are determined by $[(1 + \cos\theta)/2]^2$, $(\sin^2\theta)/2$, and $[(1 - \cos\theta)/2]^2$ respectively.[38] Therefore, in the conventional gMOT with no retroreflected beams, the proportion of the σ⁻ component is minimal. Consequently, optical pumping into $m_F = -F$ is not effective for atomic species with high $F$ owing to the presence of many Zeeman sublevels. This leads to a lack of cyclic σ⁻ transitions between $m_F = -F$ and $m_{F'} = -F'$, which are crucial for exerting position-dependent restoring forces on an atom, in $z > 0$. Therefore, the position of the cold atom cloud shifts more significantly toward the $+z$-direction as $F$ increases. Consequently, the quadrupole magnetic field, represented as $|B(x, 0, z)| \propto \sqrt{z^2 + x^2/4}$, results in a nonlinear Zeeman shift that is not conducive to restoring forces in the horizontal direction. This complexity makes it challenging to generate a conventional gMOT for alkali metal atoms with high $F$. However, a balanced gMOT can be achieved by utilizing a retroreflected beam with adjusted intensity to bring the cold atom cloud back to the center of the quadrupole magnetic field. Furthermore, the retroreflected beam with σ⁻ polarization should induce optical pumping into $m_F = -F$, facilitating the cyclic σ⁻ transitions in $z > 0$.

**Acknowledgment**

This study was supported by Innovative Science and Technology Initiative for Security Grant No. JPJ004596, ATLA, Japan. A part of this work was supported by "Advanced Research Infrastructure for Materials and Nanotechnology in Japan (ARIM)" of the Ministry of Education, Culture, Sports, Science and Technology (MEXT). Proposal Number JPMXP1223AT0240. The authors express their gratitude to T. Ikegami, formerly with the Micromachine Center, for valuable discussions.

**Conflict of Interest**

A. Takamizawa, R. Hokari, S. Kagami, Thu H. H. Le, and R. Takei have a pending patent (Japanese Patent Application No. 2024-147567).

**Author Contributions**

**A. Takamizawa**: Conceptualization (equal); Data curation (lead); Formal analysis (lead); Funding acquisition (supporting); Investigation (lead); Methodology (equal); Visualization (lead); Writing - Original draft (lead). **R. Hokari**: Funding acquisition (supporting); Investigation (supporting); Resources (lead). **S. Kagami**: Conceptualization (supporting); Funding acquisition (supporting); Investigation (supporting); Methodology (equal); Validation (lead). **T. H. H. Le**: Conceptualization (equal); Funding acquisition (supporting); Methodology (equal). **R. Takei**: Funding acquisition (supporting); Investigation (supporting); Resources (supporting). **K. Hagimoto**: Resources (supporting); Visualization (supporting). **S. Yanagimachi**: Funding acquisition (lead); Supervision (lead).

**Data availability statement**

The data supporting the findings of this study are available from the corresponding author upon reasonable request.